\documentclass[usenatbib]{mnras}

\usepackage{newtxtext,newtxmath}

\usepackage[T1]{fontenc}

\DeclareRobustCommand{\VAN}[3]{#2}
\let\VANthebibliography\thebibliography
\def\thebibliography{\DeclareRobustCommand{\VAN}[3]{##3}\VANthebibliography}

\usepackage{graphicx}%
\usepackage{multirow}%
\usepackage{amsmath}%

\usepackage{amssymb}%
\usepackage{mathrsfs}%
\usepackage[title]{appendix}%
\usepackage{xcolor}%
\usepackage{textcomp}%
\usepackage{manyfoot}%
\usepackage{booktabs}%
\usepackage{algorithm}%
\usepackage{algorithmicx}%
\usepackage{algpseudocode}%
\usepackage{listings}%
\usepackage{enumitem}
\usepackage{subcaption}
\usepackage{float}
\usepackage[section]{placeins}
\usepackage{capt-of}

\title[Machian MOND: a variable $a_0$ in galaxy clusters]{Machian MOND: a variable $a_0$ in galaxy clusters}

\author[Manuel Urueña Palomo, Juan David Santander]
        {Manuel Urueña Palomo$^{1}$\thanks{E-mail: yq87exim@studserv.uni-leipzig.de, muruep00@gmail.com},
        Juan David Santander$^{2}$\\
$^{1}$Universität Leipzig, Faculty of Physics and Earth System Sciences, Linnéstraße 5, Leipzig (04103), Germany\\
$^{2}$UNICEF$^{\dagger}$, United Nations Plaza 3, New York (10017), USA}

\pubyear{\the\year{}}

\begin{document}
\label{firstpage}
\pagerange{\pageref{firstpage}--\pageref{lastpage}}
\maketitle

\begin{abstract}
Modified Newtonian Dynamics (MOND) generally resolves the need for dark matter in galaxy rotation curves introducing a single new constant of acceleration $a_0$. It is well known that increasing $a_0$ by a factor of a few can alleviate the residual mass discrepancies that MOND leaves in galaxy clusters. Within a parameter-free Machian interpretation of MOND motivated by $a_0\sim GM_u/R_u^2$ arising from the scalar sum of inverse-square distance gravitational mass contributions in the universe, we promote $a_0$ to a variable in spherically symmetric mass distributions, which is sensitive to mass external to the locally enclosed spherical region. Instead of a boost of $a_0$ in terms of gravitational potentials as in EMOND, we show that a boost in terms of this directionless inverse-square field roughly amounts to the boost needed to accommodate the mass discrepancies of MOND in galaxy clusters, while retaining MOND predictions in galaxies. We begin to generalize this approach by defining a new $a'_0=G\sum_{i}m_i/|\mathbf{r}-\mathbf{r_i}|^2-a_N$, where $a_N$ is the magnitude of the usual Newtonian acceleration.
\end{abstract}


\section{Introduction}\label{sec1}

\subsection{MOND and galaxy clusters}

\begingroup
\renewcommand{\thefootnote}{\ensuremath{\dagger}}
\footnotetext{This work was conducted in the author’s personal capacity and does not represent the views or positions of UNICEF. jdsantander@gmail.com}
\endgroup

The observation of rotational velocities in galactic disks higher than those predicted by Newtonian gravity, as shown in galaxy rotation curves by \cite{Rubin1983}, has motivated research into modifications of the laws of gravity and mechanics as alternatives to the still elusive physical dark matter \citep{Review}. After careful observation of the patterns and regularities found in these systems related to visible mass only, such as the baryonic Tully–Fisher relation (the asymptotic rotational velocity scales as the fourth root of the baryonic mass), Renzo's rule (features in the visible mass distribution strongly correlate with features in the velocity profiles) and the empirical observation that Newton’s laws accurately predict dynamics only above a certain acceleration scale, Mordehai \cite{Milgrom1983a,Milgrom1983b,Milgrom1983c} proposed the simplest modification to Newtonian dynamics consistent with these observational constraints.

The original formulation of Milgrom's Modified Newtonian Dynamics MOND results in a constant asymptotic rotational velocity $v=\sqrt[4]{GMa_0}$ in the low acceleration or deep-MOND regime with the need to introduce an empirically determined constant with dimensions of acceleration $a_0$. The transition between the classical Newtonian regime where $v=\sqrt{GM/r}$ and the deep-MOND regime is governed by an ‘interpolating function' $\mu \left(a/a_0\right) = 1/ (1+ \left(a_0/a\right)^n)^{(1/n)}$ (with $n=1$ for the simple and $n=2$ for the standard interpolating functions) based on $|\mathbf{a}|=a$ as the measured or ‘true acceleration' and the constant $a_0$ (also referred to as the Hubble acceleration or the acceleration scale constant). Such constant sets the characteristic acceleration scale below which MOND effects manifest. The interpolating function, interpreted as modified inertia $\mathbf{F}=m_i\mu(a/a_0)\,\mathbf{a}$, must satisfy the limiting conditions $\mu(x) \rightarrow x$ for $x\ll 1$ to reproduce the Tully-Fisher law, and $\mu(x) \rightarrow 1$ for $x\gg 1$ to recover the usual Newtonian laws at high accelerations. Switching to an interpolating function $\mathbf{a}=\nu(a_0/a_N)\,\mathbf{a_N}$ in terms of the Newtonian acceleration $|\mathbf{a_N}|=a_N$, $\nu(a_0/a_N)=(1/2)^{1/n}(1+\sqrt{1+4(a_0/a_N)^n})^{1/n}$ (with $n=1$ for the simple and $n=2$ for the standard interpolating functions) and

\begin{gather}\label{eq1}
\mathbf{a_{MOND}}=\nu\left(\frac{a_0}{a_N}\right)\,\mathbf{a_N} \sim \mathbf{a_N}\sqrt{1+\frac{a_0}{a_N}},
\end{gather}

so that at the deep-MOND regime, $a\sim \sqrt{a_Na_0}$.

To satisfy conservation laws and for computation purposes, Lagrangian-based models have been developed, such as AQUAL and QUMOND. Both are special cases of three-potential (TRIMOND) theories \citep{TRIMOND} that recover Milgrom’s MOND in highly symmetric systems, while showing modest deviations elsewhere. Nevertheless, as non-relativistic models lacking a physical explanation for the origin of the interpolating function and the acceleration scale constant, MOND is best regarded as an effective theory or an approximation to a more fundamental modification of gravity or inertia. Relativistic MOND theories, such as TeVeS \citep{TEVES}, recover MOND's predictions for galaxies through extra scalar and vector fields and several accompanying parameters and constants.

Milgrom’s MOND, while remarkably successful in describing galactic rotation curves, remains insufficient to explain all phenomena attributed to dark matter, such as galaxy cluster dynamics \citep{Sanders1998,Sanders2003,Pointecouteau2005}. Roughly 90$\%$ of the visible (baryonic) mass in galaxy clusters is in the form of hot and diffuse intracluster gas, with the remainder contained in the galaxies themselves. Observations of galaxy motions, X-ray–emitting gas, weak and strong gravitational lensing, and the Sunyaev–Zel’dovich effect all indicate a persistent mass discrepancy between the total (dynamical or lensing) mass and the observed baryonic matter. Under Newtonian dynamics, this discrepancy decreases with increasing distance from the cluster core: in the central regions it is typically of order $10-50$, falling to values of $5-10$ at radii of about $1-2\,$Mpc.

There is evidence for an analogous radial acceleration relation in galaxy clusters; an empirical scaling law between the observed acceleration and the expected from visible mass only, with a characteristic acceleration scale of $\sim10a_0$ \citep{RAR}. The typical observed accelerations in the cores of clusters are on the order of, or a few times larger than $a_0$, and MOND predicts only a small correction in them, underpredicting mass discrepancies by a factor of a few \citep{Milgrom2008}, around $2-6$ \citep{Kelleher}. At intermediate radii of about $1\,$Mpc, the measured accelerations are around $0.2\,a_0-0.5\,a_0$ \citep{Sanders1998}, and MOND reduces the mass discrepancies at these radii to factors of only $2-3$ \citep{Ettori2019}. At radii above $2\,$Mpc, the accelerations drop to roughly $0.05\,a_0$, and MOND leaves an even smaller residual mass discrepancy, as low as only around $15\%$. A recent study suggests that the previously described discrepancies may be significantly alleviated \citep{Kroupa}.

In analogy with Milgrom’s identification of a characteristic acceleration scale as the basis for modifying gravity or inertia in galaxies, Bekenstein identified that a gravitational potential (or velocity) scale could play a similar role for MOND in galaxy clusters. Given that clusters have huge mass distributions that are much more spatially extended than galaxies, their gravitational potentials are correspondingly deeper, and a few times larger $a_0$ can bridge the mass discrepancy gap in clusters within MOND. In his scale-dependent critical acceleration formulation, Bekenstein introduced the required boost in galaxy clusters of $2-3$ times $a_0$ while leaving the galactic regime nearly unchanged through $a_0\,e^{-2\Phi /s^2}$ with $\Phi$ the negative gravitational potential and $s$ a data-fitted velocity scale on the order of $10^6$ m/s \citep{Bekenstein}. HongSheng Zhao and Benoit Famaey developed Extended MOND (EMOND) \citep{EMOND,EMOND2,EMOND3} in a modified AQUAL Lagrangian to generalize Bekenstein's idea and address the problematic emergence of unrealistically large effective $a_0$ values in regions of extremely high gravitational potential $\Phi\sim -c^2$ such as neutron stars and stellar black holes, where any modification to gravity is strongly constrained by General Relativity (GR). But together with an additional interpolating function, they also need to introduce a dimensionless constant parameter $\beta \sim 8\times 10^{-3}$ tweaking the gravitational potential scale of the speed of light squared to fit the data. Another extension of MOND addressing the cluster problem is GMOND \citep{GMOND}, in which a third regime for clusters is introduced as a stronger inverse-square law by a fine-tuned parameter effectively boosting the gravitational constant. More recently, a p-Laplacian generalization of AQUAL has been suggested to accommodate dwarf spheroidal galaxies and galaxy clusters by allowing the exponent $p$ to depend on the properties of the system through an extra function \citep{Scherer}.

Progress in fundamental physics has often been achieved by reducing the number of fundamental constants\footnote[1]{Such as the gravitational acceleration at the surface of the Earth and other planets or Kepler's 3rd law constant for a universal gravitational constant, the unification of the speed of light with vacuum permittivity and permeability, and atomic constants, such as Rydberg's constant, for the Planck constant.}. Therefore, given that MOND only introduces a single new constant into Newton's laws (which will be considered a derived parameter in the next section), it is of considerable theoretical interest to explore modifications to MOND without the need for more fundamental constants or free parameters, in the search for the physical origin of MOND.

\subsection{Machian MOND}

Mordehai Milgrom has insisted on several occasions that MOND might be a consequence of Mach's principle:

\begin{quote}
    
\textit{An attractive possibility is that MOND results as a non-relativistic, small-scale expression of a fundamental theory by which inertia is a vestige of the interaction of a body with ‘the rest of the Universe', in the spirit of Mach’s principle.} [...] \textit{We thus envisage inertia as resulting from the interaction of the accelerated body with some agent field, perhaps having to do with the vacuum fields, perhaps with an “inertia field” whose source is matter in the “rest of the universe”–in the spirit of Mach’s principle.} \citep{Moti1}

\end{quote}

\begin{quote}

\textit{It seems to me that in looking for an ultimate theory, the Mach principle may serve as a most useful guide.} [...] \textit{The possible connection between such numerical “coincidences”, which relate parameters of “local” physics to cosmological parameters and the Mach principle, is quite obvious.} \citep{Milgrom1983a}

\end{quote}

but has not developed this relationship further. The coincidences $a_0\sim c^2\sqrt{\Lambda}$ and $a_0\sim cH_0\sim c^2/R_u$ have gained more attention as potential evidence that MOND arises from the quantum vacuum or cosmic acceleration \citep{Milgrom2020}. In contrast to the Machian coincidence $a_0\sim GM_u/R_u^2$, where $M_u$ (the mass within the horizon of radius $R_u$) influences local dynamics as in most Machian-inspired theories, the cosmological constant $\Lambda$ remains obscure. Not only must its existence be assumed, since the evidence for it is purely indirect, but one must also assume that it corresponds to the quantum vacuum energy density driving the universe’s accelerated expansion. Following Occam's razor, $a_0(M_u,R_u)$ is more plausible, since we are certain about the existence of $M_u$ and $R_u$ by direct observations, but uncertain about $\Lambda$ and the origin of the accelerated expansion. Other authors have drawn attention to a possible Machian origin of MOND \citep{Corda2016,Corda2024,Darabi2010,Gine2009,Gine2012}.

It is not our purpose to explain here the deep rabbit hole of Mach's principle \citep{Mach1881} and the relativity of inertia, despite the historical ambiguity\footnote[2]{One of such misunderstandings is that it necessarily implies the anisotropy of inertia, which is highly constrained by observations and effectively ruled out. Not only it was simply suggested as a possible feature of Mach's principle and not as a consequence by Cocconi and Salpeter \citep{Cocconi1958}, but it was argued by Robert Dicke \citep{Dicke} that such anisotropic effects would be undetectable in a metric theory. The anisotropy is also avoided if inertial mass is a scalar.} surrounding their interpretation, even though such an understanding is required to realize its connection to MOND. An introduction to the Machian reasoning relevant to MOND has already been developed by \cite{Manuel}, and Mach's principle has been extensively covered by \cite{Reinhardt,Barbour,Bondi1997}. One of the expected consequences\footnote[3]{The other two expected consequences correspond to the Lense-Thirring effect and to the fact that a body should experience an acceleration if nearby masses are accelerated, with the same direction as their acceleration. Both are featured in GR by frame-dragging.} of Mach's principle according to \cite{Einstein1922} is that “\textit{The inertial mass of a body should increase with the agglomeration of masses in its neighborhood}'', or equivalently, \textit{the inertia of a body should decrease when masses are removed from its neighborhood}. Such consequence, unlike the other two Einstein mentioned, is not satisfied by GR, in which inertia is still absolute\footnote[4]{Flat Minkowski spacetime is a solution to GR in absence of masses in which inertia behaves as usual. It contradicts Mach's principle since inertia should have exclusively an origin in the mass distribution at large.}. In MOND, this isolation of a body from other masses trivially corresponds to the deep-MOND regime, where the system is subjected to very low gravitational field intensities.

Given that the gravitational potential of the universe $GM_u/R_u \sim c^2$ does not naturally correspond to the potential scale featured in EMOND removing MOND's problem in galaxy clusters, and that a huge dimensionless correction factor (as a dimensionless parameter) must be introduced to adjust this scale, we instead turn to local field intensities as the ones relevant for modifying $a_0$ in an attempt to address the cluster problem. We are motivated by the notion that since accelerations in clusters are of the order of $a_0$, if these sum to $a_0$ through some as yet unknown mechanism in clusters, the required boost to $a_0$ could be approximately achieved without introducing new constants.

\section{Variable $\lowercase{{a_0}}$ in Machian MOND}\label{sec2}

\subsection{Mach against Newton's shell and Birkhoff's theorems}

A brief exception to omitting the history of Mach’s principle is made here as a justification for the physical foundations of the following section.

\begin{figure}
\centering
\includegraphics[scale=0.105]{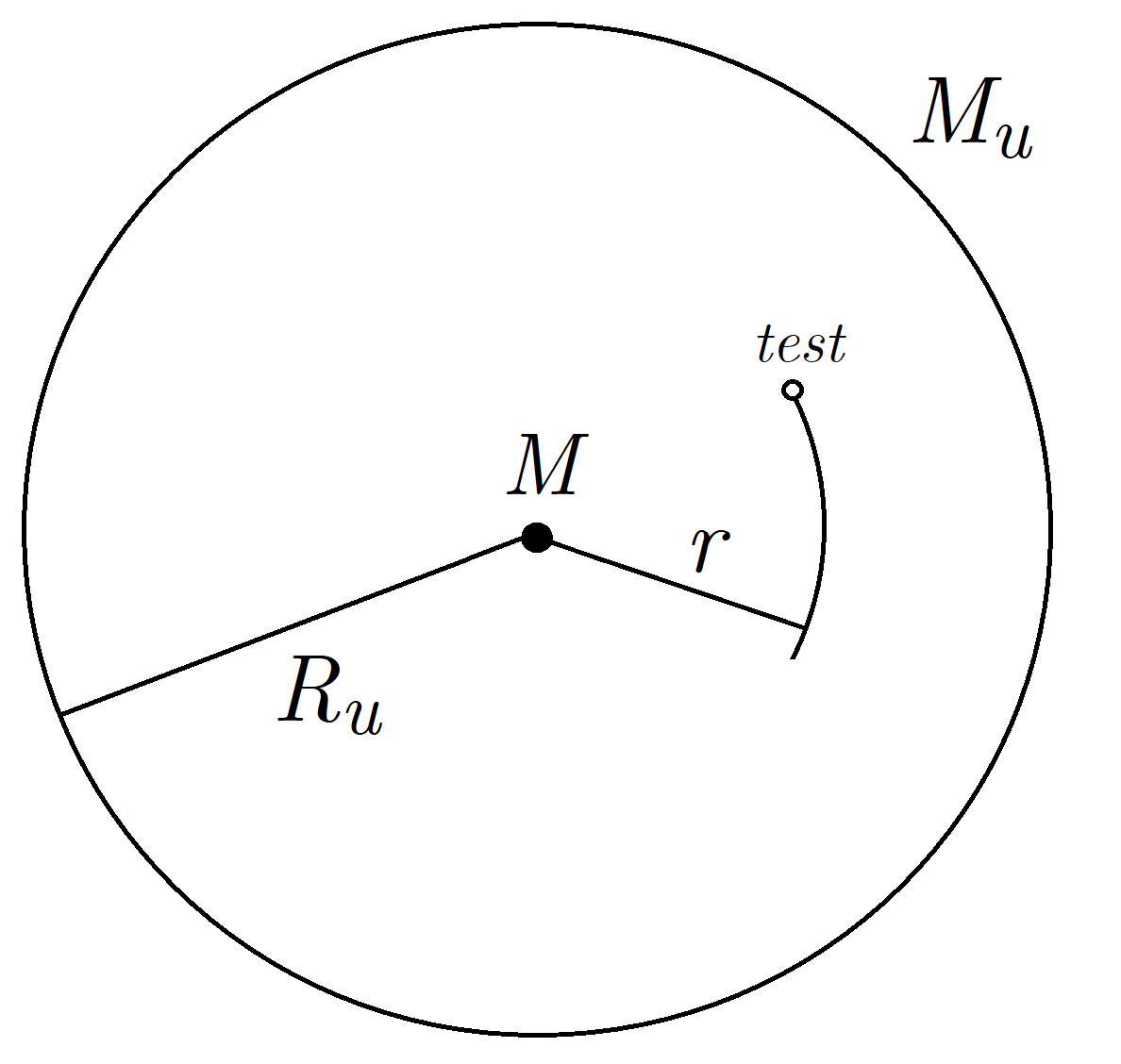}
\caption{Brans’ shell model of the universe: the total mass $M_u$ within the observable universe is represented by an equivalent static mass shell of radius $R_u$.}
\label{Fig1}
\end{figure}

At the beginning of the `golden age' of GR, \cite{Brans} presented a thought experiment considering the ratio between the inertial and gravitational masses of a body within a Machian framework. He idealized a static universe consisting solely of a massive hollow shell of radius $R_u$ and inertial mass $M_u$ (serving as a simplification to the large-scale mass and size of the universe, with approximately equivalent potential at its center), together with a relatively small body of inertial mass $M$ at the center and a test particle (Figure \ref{Fig1}). 

He expected the acceleration of the test particle to be independent of its own mass (as required by the equivalence principle), to depend on the mass $M$ and its distance $r$ from $M$, and because of Machian arguments about the inertial properties of the test body being determined by the distribution of matter in the universe and not being intrinsic of the body, conceivably also to depend on $M_u$ and $R_u$. Thus, $a=f(M_u,R_u)M/r^2$ and he related $f$ to the gravitational constant by $f=G\sim c^2/(M_u/R_u)$ recovering the Newtonian gravitational acceleration. This relationship for the gravitational constant\footnote[5]{By dimensional analysis, the combinations $G\sim c^2/(M_u/R_u)$ and $a_0\sim GM_u/R_u^2$ are essentially the only ways (up to additional dimensional constants) to construct constants that involve both a characteristic mass and distance, thereby encoding properties of a mass distribution.}, which inspired the Brans-Dicke theory, dates back to \cite{Sciama1953} with its roots in the works of \cite{Schrodinger1925} and \cite{Reissner1915} on Mach's principle. The difference in predictions between the Brans-Dicke theory and GR has been locally tested in the Solar System near the Sun through the Shapiro time delay effect measured by the Cassini experiment, showing that the Brans-Dicke dimensionless coupling factor of the scalar field to gravity, which differentiates between the theory and GR, must be so high that Brans-Dicke theory is almost indistinguishable from GR. All of these tests have been performed in the high-acceleration regime of the Solar System.

Milgrom's MOND can be formulated very similarly to Brans’ Machian idea with $a=f(M_u,R_u,M,r)M/r^2$ by introducing\footnote[6]{We also introduce the Machian gravitational constant defined in terms of the same parameters $G\sim c^2/(M_u/R_u)$, although not strictly necessary. Within the interpolating function, the gravitational constant cancels out, remaining only in the usual Newtonian force law.} $a_0\sim GM_u/R_u^2$ as

\begin{gather}\label{eq2}
a_{\mathrm{Machian \, MOND}}\sim c^2\frac{M/r^2}{M_u/R_u}\sqrt{1+\frac{M_u/{R_u}^2}{M/r^2}}.
\end{gather}

With the interpolating function acting on Newton’s second law rather than modifying gravity, the resulting inertial law depends on the inertial mass of the test particle, the gravitationally active mass $M$ and the mass of the rest of the universe $M_u$, i.e., inertia becomes a function of all relevant mass distributions, in the spirit of Mach.

Brans' thought experiment illustrates how Mach's principle violates Newton's shell theorem\footnote[7]{The gravitational potential inside an empty spherically symmetric massive shell is constant with vanishing field intensities. Consequently, a body inside such shell experiences no net gravitational force from the shell, regardless of the mass of the body or the shell, or the position within the shell.}. According to Newton, a spherically symmetric massive shell does not contribute to the local dynamics of bodies within it when interior masses are present. In Brans’ framework the acceleration depends on the shell’s mass and radius.

A consequence of the shell theorem is that the gravitational field vanishes at the center of any spherically symmetric mass distribution. For any event, there exists a surrounding spherical region of the observable universe with approximately uniform matter density, for which the net gravitational field due to the mass of the universe vanishes. Therefore, the term $M_u/R_u^2$ in (\ref{eq2}) (for which $a_0$ is effectively replaced) is not a field intensity or acceleration, but represents the directionless sum of the inverse-squared mass contributions from all volume elements in the universe at large.

\begin{gather}\label{eq3}
a_{0} =G\int_{u} \frac{dm}{r^2} = 4\pi G\int_0^{R_u} \rho_u \, dr=\frac{3GM_u}{R_u^2}.
\end{gather}

This $a_0(M_u,R_u)$ scalar term matching $a_0=(1.2\pm 0.3 )\times 10^{-10}$ m/s$^2$ \citep{EMOND} will be used in the following sections as a constant at present epoch and simply denoted as $a_0$.

Newton's shell theorem follows from the inverse-square law. MOND breaks with the shell theorem in the MOND regime except for spherical symmetry\footnote[8]{For spherically symmetric systems in the deep-MOND regime, the gravitational field is exactly zero at the very center; outside the spherically symmetric mass distribution, the gravitational field is the same as if all the mass were concentrated at the center.}, since the effective deep-MOND force tends to decay with the inverse of the distance at low accelerations, and superposition does not hold. Bekenstein's scale dependent critical acceleration in AQUAL and its generalization in EMOND based on a modification to $a_0$ in terms of potentials, both break with the shell theorem even in spherically symmetric cases, and masses outside a sphere can affect local dynamics inside it. This effect might even help with situations such as the Bullet cluster, in which the sources of the gravitational field seem to be outside the region expected to govern local dynamics in a Newtonian fashion. The external field effect (EFE) in Milgrom's MOND is also an example of how external far away mass distributions influence local dynamics in MOND.

A generalization of Newton's shell theorem occurs in GR for test particles inside a massive shell by Birkhoff's theorem, where they experience no accelerations from the shell. For off-centered massive particles, their presence inside the shell breaks the exact symmetry of spacetime, leading to weak accelerations toward the center of the shell. A stronger effect is found in AQUAL \citep{Birkhoffs}.

By identifying MOND's acceleration scale constant as coming from Brans' shell model of the universe $a_0\sim GM_u/R_u^2$, we will argue in the following that $a_0$ is modified in clusters.

\subsection{Machian MOND and a variable $a_0$}\label{section}

Machian MOND (\ref{eq2}) is built under the assumption that $a_0\sim GM_u/R_u^2$ while local gravitational field intensities are only accounted for in $a_N= GM/r^2$. This formulation raises the question of when a mass distribution should contribute to local field intensities and when it should instead contribute to $a_0$, and why Machian MOND appears to distinguish so clearly between the two.

\begin{figure}
\centering
\includegraphics[scale=0.115]{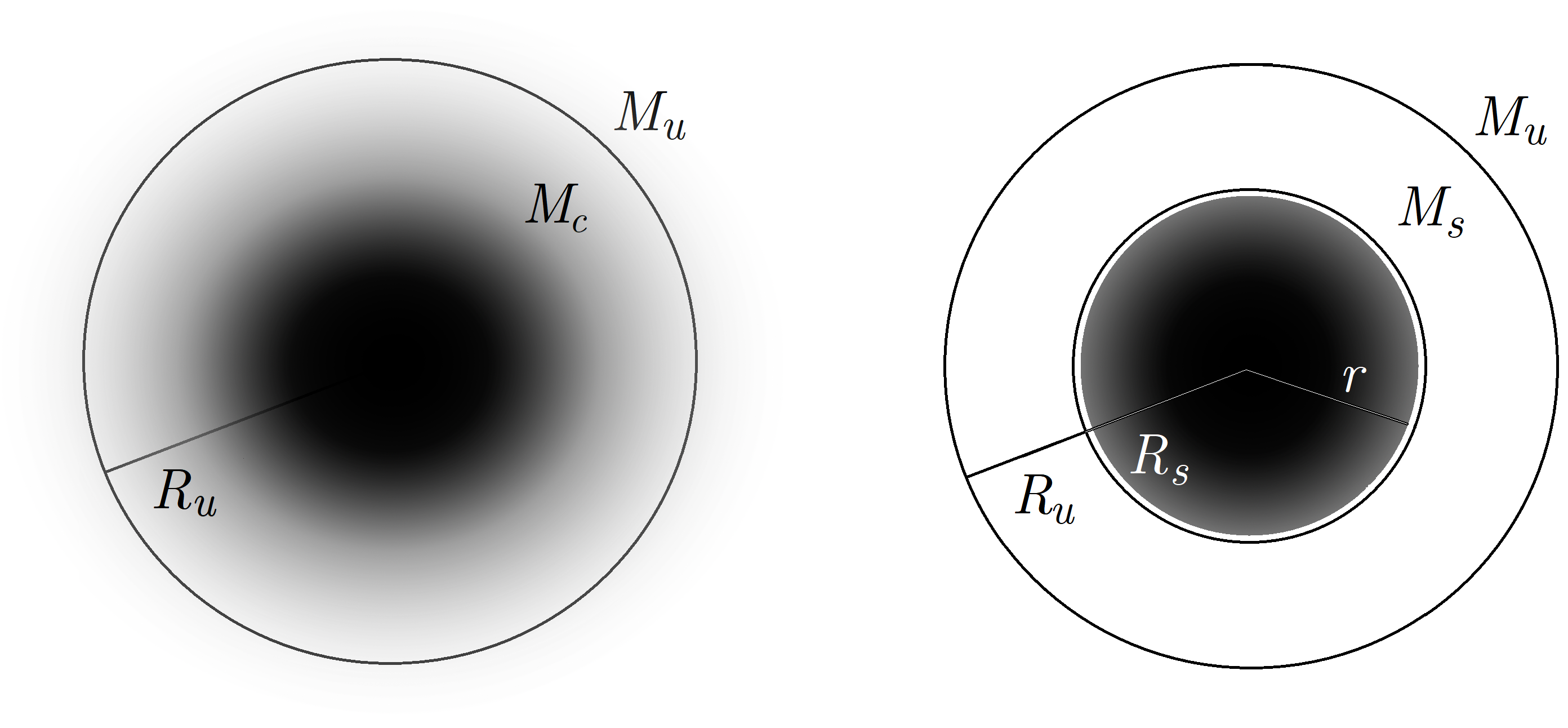}
\caption{Left: a spherically symmetric schematic model for a galaxy cluster mass distribution (blurred) with total mass $M_c$ inside Brans' shell of the universe $M_u$ and $R_u$. Right: the mass distribution of the cluster exterior to a radius $r$ is idealized as an additional shell with mass $M_s$ and radius $R_s$. Here $R_s\neq r$ depends on the mass distribution outside $r$. Not to scale.}
\label{Fig2}
\end{figure}

Subsequently, it is conceivable that perhaps $a_0$ is also affected by large but local mass distributions analogous to that of the universe, such as in galaxy clusters.  If the origin of $a_0$ is idealized as arising from an exterior spherically symmetric shell of mass $M_u$ at distance $R_u$ of the universe, such as Brans' model, then a galaxy within a galaxy cluster may likewise be regarded as being enclosed by an additional symmetric shell of mass $M_s$ at a radius $R_s$, representing the cluster mass exterior to the enclosed mass within radius $r$ of that galaxy (Figure \ref{Fig2}).

We denote $a_s\sim GM_s/R_s^2$ the gravitational scalar sum of the inverse-squared masses of the cluster exterior to the radius $r$ and expect a non-linear dependence of the effective acceleration scale constant $a'_0(M_u,R_u,M_s,R_s)$. We propose the following function as an Ansatz:

\begin{gather}\label{eq4}
\mathbf{a}=\mathbf{a_N}\sqrt{1+\frac{a_0}{a_N}}\sqrt{1+\frac{a_s}{a_N}} \ , \quad a=\sqrt{(a_0+a_N)(a_s+a_N)},
\end{gather}

which is equivalent to Milgrom's MOND but replacing the acceleration scale constant $a_0 \rightarrow a'_0=a_0+a_s+a_0a_s/a_N$. Thus, $a'_0$ is no longer a fundamental parameter, but an effective, position-dependent quantity arising from the mass distribution. Note that $a_N$ is an input field determined by the mass distribution, independent of the test particle motion or its dynamics. It satisfies the following desired conditions of an extension for MOND:

a) It does not introduce new constants or adjusted parameters, as required in other proposals such as Bekenstein's scale dependent critical acceleration or EMOND.

b) It reduces to Milgrom's MOND when the newly introduced $a_s\rightarrow 0$, i.e., when there is no external local shell. The EFE can be approximated as usual through $a_N$.

c) Removing the mass of the distant universe $M_u\rightarrow 0$, $a_0\rightarrow 0$, $\mathbf{a}=\mathbf{a_N}\sqrt{1+\frac{a_s}{a_N}}$ and the shell of the cluster behaves as a new `cosmic' shell, playing the same role in the dynamics as $a_0$ does in Machian MOND.

If $a_N \gg a_0, a_s$ then $a \approx a_N$ and the system is in the Newtonian regime. If $a_N \ll a_0$ and $a_s$ is negligible, then $a \approx \sqrt{a_N a_0}$, corresponding to the usual deep-MOND regime. When $a_N \sim a_s \sim a_0$, equation (\ref{eq4}) accounts for the combined contributions of both the cosmic and local external shells. In the following section, a simple toy model will show that such a modification to MOND produces a boost to $a_0$ in galaxy clusters of the order required by observations.

\subsection{A toy model for galaxy clusters}

We test the proposed modification to Milgrom's MOND (\ref{eq4}) with a simple spherically symmetric model of standard galaxy clusters, in which $a_s(r)$ and $a_N(r)$ vary with distance from the cluster core $r$. Our aim is to show that such a modification generally matches the required boost in $a_0$ to fit clusters, while leaving MOND predictions for individual galaxies effectively unchanged, as discussed in the next section.

For any given distance $r$ from the core of the cluster, the cluster's mass $M_s$ exterior to $r$ is treated as an additional shell, analogous to Brans' cosmic shell. Its field contribution $a_s(r)$ is computed\footnote[9]{We used a simplified equation for (\ref{eq5}) as $a_s(r)=G\int_r^R dm/r'^2$ in the published version, integrating the mass density divided by the distance to the center of the cluster $r'^2$ instead of to the point of evaluation $|r-r'|^2$. While not significantly altering the results in comparison to those presented here, we thank Hongsheng Zhao for suggesting this change.} analogously to the definition of $a_0$: the scalar sum of the inverse-squared mass contributions

\begin{gather}\label{eq5}
a_{\mathrm{s}}(r)= G\int_{r<r'<R} \frac{\rho(r')}{|\mathbf{r}-\mathbf{r'}|^2}d^3r' = \frac{2\pi G}{r} \int_{r}^{R} \rho(r') r' \ln\left( \frac{r+r'}{r'-r} \right) dr',
\end{gather}

with $R$ the virial radius, $\rho(r')$ the baryonic mass density of the cluster, so that $a_s\sim GM_s/R_s^2$. The virial radius marks the boundary of virial equilibrium. Although our models consider only the baryonic component of the cluster, we make use of the standard dark-matter-based virial radius, which serves as a convenient cutoff for integral and the definition of the cluster’s total mass. Considering a larger bound than $R$ in the integral leads to a very slight boost of $a_s(r)$ due to the $1/r^2$ dependence and the fact that those additional outer masses of the cluster lie very far from the center. For instance, for a greater bound of $4R$, the boost to $a_s(r)$ remains below $25\%$ at the core of one of the toy models and only mildly increase $a_s(r)$ far away from the core (Figure \ref{Fig6}). For even larger bounds, $a_s(r)$ converges, and such a boost can be considered contained in $a_0$ (\ref{eq3}), since a boost to $a_s(r)$ is equivalent to a boost to $a_0$.

We choose to model a Coma-like, Virgo-like, and Fornax-like type clusters using a standard spherically symmetric $\beta$-model density profiles of baryonic matter $\rho(r) = \rho_0 \left[ 1 + \left( r/r_c \right)^2 \right]^{-3\beta/2}$ with interior enclosed mass $M(< r) = \int_0^r 4 \pi r'^2 \rho(r') \, dr'$ and exterior mass $M(> r) = \int_r^R 4 \pi r'^2 \rho(r') \, dr'$, Newtonian acceleration $a_N(r) = G M(<r)/r^2$, and $a_{\mathrm{MOND}} = \sqrt{a_Na_0}$ with $a_0=1.2\times 10^{-10}$ m/s$^2$ as an approximation. While a  multi-component mass model would be more accurate, a single standard $\beta$-model offers a good match for emission profiles of non-cool-core clusters with a flat surface brightness profile in their cores, such as the Coma cluster \citep{Eckert2012}. For the cool-core clusters of Virgo and Fornax, a single $\beta$-model underestimates central density at $r<300\,$kpc and $r<100\,$kpc respectively, and their models serve as an order-of-magnitude consistency check only.  

The adopted parameter values (Table \ref{Table1}) are chosen to be representative of low--redshift clusters and to lie within the observational ranges reported in large X--ray surveys of nearby systems. The parameter choices for a Coma-like cluster are guided by \cite{Briel1992} and typical intervals are $\beta \sim 0.60-0.80$, core radius $r_c \sim 0.25-0.40 \,$Mpc, central gas densities $\rho_0 \sim (5-7) \times 10^{-24}$ kg/m$^{-3}$, and virial radii $R \sim2.6-3.0 \,$Mpc. For a Virgo-like cluster following \cite{Schindler1999}, $\beta \sim 0.45-0.50$, $r_c \sim 0.05-0.1 \,$Mpc, $\rho_0 \sim (2-4) \times 10^{-24}$ kg/m$^{-3}$, and $R \sim0.8-1.2 \,$Mpc. For a Fornax-like cluster \citep{Machacek2005}, $\beta \sim 0.40-0.55$, $r_c \sim 0.04-0.08 \,$Mpc, $\rho_0 \sim (2-4) \times 10^{-24}$ kg/m$^{-3}$, and $R \sim0.5-1.0 \,$Mpc. The following selected values used for each cluster type fall within these envelopes and are intended to represent Coma--like, Virgo--like, and Fornax--like systems rather than their exact replicas:

\begin{center}
\captionof{table}{Parameters for the toy models of galaxy clusters.}
\label{Table1}

\resizebox{\columnwidth}{!}{%
\begin{tabular}{cccccc}
\toprule
Cluster type & $\beta$ & $r_c$ & $\rho_0$ & $R$ & Baryonic mass \\
             & [$-$] & [kpc] & [$10^{-24}\,\mathrm{kg\,m^{-3}}$] & [Mpc] & [$10^{14}M_\odot$] \\
\midrule
Coma--like   & 0.75 & 350 & 5.00 & 2.80 & 1.80 \\
Virgo--like  & 0.50 & 80  & 3.75 & 1.00 & 0.10 \\
Fornax--like & 0.45 & 50  & 3.00 & 0.80 & 0.04 \\
\bottomrule
\end{tabular}%
}
\end{center}

Across all three cluster type models, the proposed formulation (\ref{eq4}) predicts accelerations that are notably higher than $a_{\mathrm{MOND}}$. Near their core, we generally find a boost in $a/a_{\mathrm{MOND}}$ of $\sim 3$ with an effective $a'_0 \sim 10\,a_0$ at a distance of $0.1$ Mpc from the center. At intermediate radii, we subsequently find a boost in $a/a_{\mathrm{MOND}}$ of $\sim 1.5$ with an effective $a'_0 \sim 2\,a_0$. The estimated true acceleration reduces to $a_{\mathrm{MOND}}$ at their virial radius, as expected from the diminishing contribution of the cluster’s outer shell mass with increasing distance from the core. Detailed data including $a/a_{\mathrm{MOND}}$ for different radii are provided as tables and plots in Appendix \ref{appendix:a}.

\section{Discussion}\label{sec3}

The model described above essentially relies on the same characteristic that gives galaxy clusters their large gravitational potential depths (and inspired modifications to MOND in terms of these), which is their very large total mass distributed over an extended, relatively uniform volume. However, in contrast to potential-based modifications such as Bekenstein's scale-dependent critical acceleration and EMOND, our model addresses Milgrom’s MOND cluster problem in terms of Newtonian accelerations and a directionless sum of inverse-square field intensities. While the EFE in MOND from an imposed external acceleration tends to ‘Newtonianize’ motion, our outer field dependence of $a_0$ can only enhance the MONDian effect (given that no shielding is possible).

We reproduce the required distance-from-the-core dependence of the MOND enhancement: strongest in the core and gradually diminishing with distance. In the central regions, the boost is somewhat smaller than needed, with ‘true’ accelerations on the order of $a_0$, rather than several times $a_0$. This may reflect the simplicity of the adopted density model for two reasons. Firstly, the core enhancement is particularly sensitive to the density–profile parameter $\beta$ in our model, as well as the total baryonic mass estimation, and we use a very simple model for the entire cluster that is not accurate enough to describe the core. Secondly, we have assumed the simplest form of the interpolating function for the contribution of $a_s$, which could actually be steeper at the cores (the same happens for the chosen MOND interpolating function (\ref{eq1}), which is required to be steeper to satisfy constraints in the Solar System). One may argue that an arbitrary complex function of measurable quantities is as unsatisfactory as a simpler function containing free parameters. However, since the required function appears to be already present in Milgrom's MOND in a similar form, we do not consider this to be a significant issue.

Given that the outer shell of clusters modifies the motion of galaxies inside clusters, we check whether such an effect is significant in the internal dynamics of galaxies themselves. This would entail that our model could also enhance rotation curves at low-acceleration regimes if $a_s\sim a_N$. In spiral galaxies, we consider a possible gentle outer shell effect from the galaxy itself because of their highly concentrated baryonic mass distribution. For a razor-thin exponential disk of the spiral Milky Way-like galaxy, $a_N\geq a_s$ holds for all regimes, and the cross-term $a_0a_s/a_N$ is negligible. For inner radius $r\sim 5\,$kpc corresponding to the Newtonian transition regime in Milgrom's MOND, $a'_0<1.5a_0$ and an effective boost to $a_0'$ of $<50\%$ is acceptable \citep{EMOND}. Such a boost can be accommodated by the uncertainty between the different proposed interpolating functions, the stellar population mass-to-light uncertainty estimations of about $10\%$, and by the uncertainty of $a_0=\left(1.2^{+0.5}_{-0.3} \right)\times 10^{-10}$ m/s$^2$ \citep{EMOND}. For smaller radii, a stronger transition of the interpolating function can accommodate any deviations from Newtonian predictions except for the singularity at the very center. For mid and larger radii $r\geq10\,$kpc, $a'_0<1.1a_0$ and the boost can easily be accommodated by the uncertainties mentioned above. For $r\rightarrow \infty$, MOND is effectively recovered.

But the shell effect from the cluster still boosts the effective $a'_0$ inside galaxies. Considering a rough estimate for the EFE as a sum of $a_N$ of the cluster and $a_N$ of the galaxy, we find that the effective $a'_0$ for a rotation curve of a galaxy is only sensitive to the cluster's $a_s$ at the core of the cluster, and only at the outer galactic radii of the rotation curve. Such an effect is also found in Bekenstein's scale dependent acceleration and EMOND, and recent studies suggest deviations from standard MOND predictions in galaxies at the cores of clusters \citep{Bilek2026}. However, in contrast to these approaches, the effective boost of $a'_0$ in our framework is partly suppressed by both the Newtonian acceleration $a_N$ of the galaxy and the cluster when considering the internal dynamics of galaxies, whereas only the cluster’s $a_N$ suppresses $a'_0$ in the dynamics of galaxies within the cluster. This effect could nevertheless help MOND in galaxies with more evenly distributed baryonic mass profiles, such as UDGs \citep{Milgrom2015}.

We also acknowledge that the concept of mass exterior to a certain radius, differentiating between $a_N$ and the cluster’s shell $a_s$, is not well-defined except for the spherically symmetric case, even within an integral formulation. It must be replaced by another more general concept that plays the same role in producing an effectively varying $a'_0$. To address this issue, we may generalize (\ref{eq4}) by interpreting that $a_s$ only comes from the external shell because it is the total directionless sum of inverse-squared distance local mass contributions minus the modulus of the Newtonian acceleration. Introducing a new scalar matter field 

\begin{gather}\label{eq6}
\phi(\mathbf{r})=\int \frac{\rho(\mathbf{r'})}{|\mathbf{r}-\mathbf{r'}|^2}d^3r',       
\end{gather}

$a_{\phi}=G\phi(\mathbf{r})$ such that $a_{\phi} \ge a_N$ and assuming linearity $a'_0=a_0+a_s$, then $a'_0= a_{\phi} - a_N$\footnote[10]{Far outside an spherical mass distribution in the high acceleration regime, $a_{\phi}\approx a_N+a_0 \sim a_N$. At the center of a uniform solid sphere in the high acceleration regime, $a_{\phi}=3GM/R^2 +a_0\sim3GM/R^2$ while $a_N=0$. At any point inside a spherical mass distribution at the high acceleration regime, $a_{\phi} \ge a_N$ while they approach each other further away from the source.} and MOND is generalized for all regimes as

\begin{gather}\label{eq7}
\mathbf{a}\sim\mathbf{a_N}\sqrt{1+\frac{a_{\phi}-a_N}{a_N}}\ , \quad a\sim\sqrt{a_Na_{\phi}} \,,
\end{gather}

where $a_{\phi}$ is the inverse-squared distance sum of total directionless field intensities $a_{\phi}\sim a_0+a_s+a_N$, similar to $a_N$ but without vector cancellations. This formulation also satisfies the desired conditions for an extension of MOND listed in \textsection~\ref{section}. For the Solar System, $a_{\phi} \sim a_N +a_0$ and we recover MOND's Newtonian regime, except for a possible but negligible boost to $a'_0$ due to the cancellation of gravitational accelerations between planets. For galaxies, the more the difference between $a_N$ and $a_{\phi}$ (i.e., the more cancellation of Newtonian fields), the bigger the departure expected from the Newtonian regime.

\section{Conclusion}\label{sec4}

Milgrom's MOND, while resolving the need for physical dark matter in galaxies, falls short in explaining the mass discrepancies found in galaxy clusters, showing more disagreement at their cores than in their outer regions. 

Machian MOND is not merely a reinterpretation of Milgrom's MOND with $a_0\sim GM_u/R_u^2$, but it is also expected to feature a variable $a_0$ in response to local mass distributions, rather than being exclusively determined by the scalar sum of inverse-squared distance mass contributions of the universe at large. The term $M_u/R_u^2$ does not represent a field intensity or an acceleration, both of which vanish at the center of any spherically symmetric mass distribution, as the observable universe is to good approximation.

In analogy with Carl Brans' shell model of the universe with mass $M_u$ and radius $R_u$, we idealize the spherically symmetric mass distribution in galaxy clusters exterior to a given radius from the core as a shell of $M_s$ and $R_s$ additional to the cosmic shell $M_u$ and $R_u$. We then consider a variable $a_0$ in terms of the cluster's exterior shell $a_s\sim GM_s/R_s^2$, using the same interpolating function of Milgrom's MOND in (\ref{eq4}), with an effective $a'_0=a_0+a_s+a_0a_s/a_N$. This parameter-free modification to Milgrom's MOND produces boosts of the same order as those typically required to resolve the mass discrepancies that MOND leaves across the cores and mid regions of galaxy clusters, depicted by simple spherically symmetric models of some clusters implementing $a'_0$. The result is a boost of $a'_0\sim 10\,a_0$ at the galaxy cluster cores, approaching standard MOND in the outer regions, and barely modifying Milgrom's MOND $a_0$ in galaxies except for the ones in the cluster's core and only at their outer radii. This matches the expected ten times larger $a'_0\sim10a_0$ required in the cores of clusters, and particularly in the Coma cluster, the several tens times $a'_0\sim30a_0$ \citep{Milgrom2015}.

Even though we do not specify a precise mechanism for how shells affect local dynamics, we start to generalize the shell formulation by introducing a scalar field $\phi=\sum_{i}m_i/|\mathbf{r}-\mathbf{r_i}|^2$ as the directionless sum of inverse-squared distance mass contributions, $a_{\phi}=G\phi$, and set $a'_0=a_{\phi}-a_N$ in MOND's interpolating function converting from Newtonian accelerations. Then MOND is simply $a\sim \sqrt{a_Na_{\phi}}$ for all regimes. The field $a_{\phi}$ gives $3GM_u/R_u^2 \sim a_0$ when evaluated on the universe at large, and $\sim 3GM_u/R_u^2+a_s+a_N$ when evaluated inside a local spherically symmetric mass distribution within the universe, such as inside a galaxy cluster. Under this generalization, the enhancement of $a'_0$, which increases with greater directional cancellation of Newtonian field intensities, is smaller than that presented in the Appendix due to dropping nonlinearity of the cross-term $a_0a_s/a_N$ of $a'_0$ in (\ref{eq4}). It is interesting to note that while the scalar field $\Phi=\sum_{i}m_i/|\mathbf{r}-\mathbf{r_i}|$ defines the gravitational constant $G=c^2/\Phi$ in Machian theories such as that of Reissner, Schrödinger, Sciama and \cite{Treder1972}, the field $\phi=\sum_{i}m_i/|\mathbf{r}-\mathbf{r_i}|^2$ defines the acceleration scale constant $a'_0=G\phi-a_N$ in Machian MOND. Especially when the deep-MOND limit is controlled only by $\mathcal{A}_0=Ga_0$. At cluster scales, a direct test for the proposed model would be to search for an environmental dependence of the effective MOND acceleration scale in galaxies located in different cluster environments. An Earth-based experiment boosting $a'_0$ appears impractical given the limited precision of current measurements of the gravitational constant, as it would require a Cavendish torsion balance inside a hollow sphere weighing over tens of tonnes.

In contrast to the extensions of MOND based on gravitational potential depths, our boost on scalar field intensities does not require tweaking to avoid high boosts at deep potential regimes such as neutron stars. This framework can set the basis for a new Lagrangian-based formulation of extended MOND with varying $a_0$ not based on gravitational potential depths.

It would be interesting to investigate if $a_0$ evolves over cosmological time, given that its definition depends on the mass enclosed $M_u$ within the radius $R_u$ of the observable universe, in a cosmological model without physical dark matter.

\section*{Acknowledgments}

We thank Hongsheng Zhao for comments on properly defining equation (\ref{eq5}).

\bibliographystyle{mnras}
\bibliography{bibliography}

\clearpage
\onecolumn

\begin{appendices}

\section{Data Tables and Plots}
\label{appendix:a}

\begin{center}
\captionof{table}{Coma type cluster: radius in [Mpc], baryonic mass in [$10^{12} M_\odot$], accelerations in [$10^{-10}$ m/s$^2$]}
\label{tab:coma}
\begin{tabular}{cccccccc}
\toprule
Radius & $M(< r)$ & $M(> r)$ & $a_N$ & $a_s$ & $a'_{0}$ & $a \,(\ref{eq4})$ & $a /a_{\mathrm{MOND}}$ \\
\midrule
0.05 & 0.04 & 179.96 & 0.02 & 0.54 & 32.63 & 0.83 & 5.22 \\
0.10 & 0.29 & 179.71 & 0.04 & 0.49 & 16.48 & 0.81 & 3.71 \\
0.25 & 3.56 & 176.44 & 0.08 & 0.36 & 7.02 & 0.75 & 2.43 \\
0.50 & 16.80 & 163.20 & 0.09 & 0.21 & 4.10 & 0.63 & 1.87 \\
0.75 & 34.43 & 145.57 & 0.09 & 0.13 & 3.18 & 0.53 & 1.65 \\
1.00 & 53.16 & 126.84 & 0.07 & 0.09 & 2.72 & 0.46 & 1.53 \\
1.25 & 71.93 & 108.07 & 0.06 & 0.06 & 2.42 & 0.40 & 1.44 \\
1.50 & 90.39 & 89.61 & 0.06 & 0.04 & 2.19 & 0.35 & 1.37 \\
1.75 & 108.45 & 71.55 & 0.05 & 0.03 & 2.00 & 0.32 & 1.31 \\
2.00 & 126.09 & 53.91 & 0.04 & 0.02 & 1.83 & 0.29 & 1.25 \\
2.25 & 143.33 & 36.67 & 0.04 & 0.01 & 1.66 & 0.26 & 1.19 \\
2.50 & 160.21 & 19.79 & 0.04 & 0.00 & 1.48 & 0.23 & 1.12 \\
2.75 & 176.73 & 3.27 & 0.03 & 0.00 & 1.26 & 0.21 & 1.04 \\
2.80 & 179.99 & 0.01 & 0.03 & 0.00 & 1.20 & 0.20 & 1.00 \\
\bottomrule
\end{tabular}
\end{center}

\begin{center}
\centering
\includegraphics[scale=0.6]{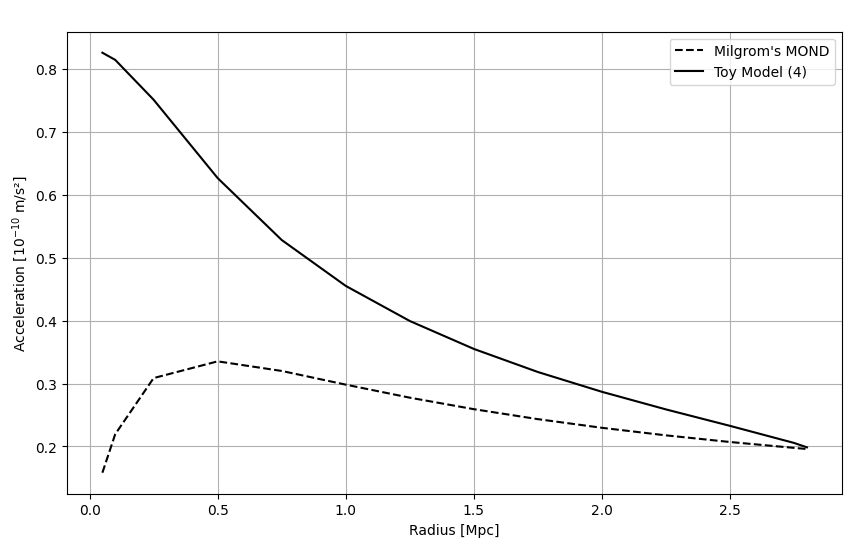}
\captionof{figure}{Plot for the Coma cluster toy model showing the acceleration predicted by MOND $a_{\mathrm{MOND}} \, (\ref{eq1})$ and by the proposed $a \,(\ref{eq4})$. }
\label{Fig3}
\end{center}

\

\

\newpage

\begin{center}
\captionof{table}{Virgo type cluster: radius in [Mpc], baryonic mass in [$10^{12} M_\odot$], accelerations in [$10^{-10}$ m/s$^2]$}
\centering
\begin{tabular}{cccccccc}
\toprule
Radius & $M(< r)$ & $M(> r)$ & $a_N$ & $a_s$ & $a'_{0}$ & $a \,(\ref{eq4})$ & $a/a_{\mathrm{MOND}}$ \\
\midrule
0.05 & 0.02 & 9.98 & 0.01 & 0.13 & 12.40 & 0.41 & 3.22 \\
0.10 & 0.15 & 9.85 & 0.02 & 0.10 & 7.12 & 0.38 & 2.44 \\
0.25 & 0.98 & 9.02 & 0.02 & 0.05 & 4.17 & 0.30 & 1.87 \\
0.50 & 3.29 & 6.71 & 0.02 & 0.03 & 2.87 & 0.23 & 1.55 \\
0.75 & 6.35 & 3.65 & 0.02 & 0.01 & 2.07 & 0.18 & 1.32 \\
1.00 & 10.00 & 0.01 & 0.01 & 0.00 & 1.20 & 0.13 & 1.00 \\
\bottomrule
\end{tabular}
\end{center}

\begin{center}
\centering
\includegraphics[scale=0.6]{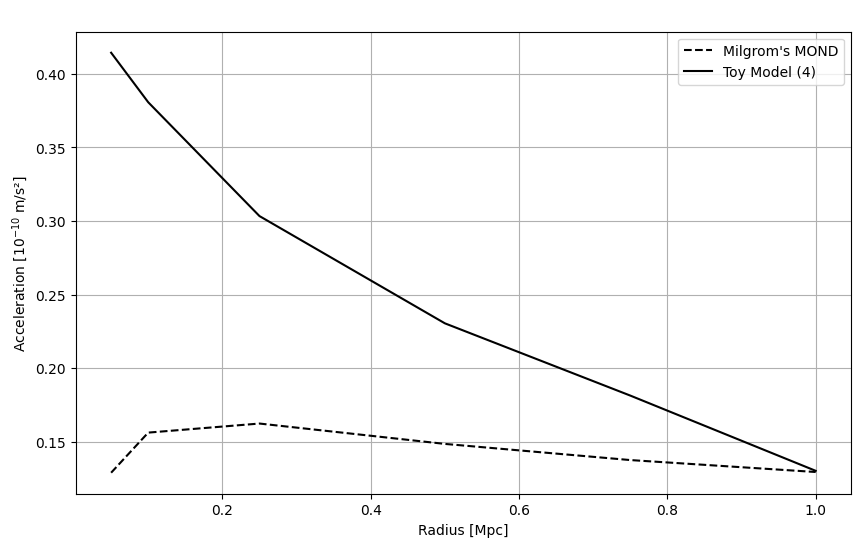}
\captionof{figure}{Plot for the Virgo cluster toy model showing the acceleration predicted by MOND $a_{\mathrm{MOND}} \, (\ref{eq1})$ and by the proposed $a \,(\ref{eq4})$. }
\label{Fig4}
\end{center}

\

\newpage

\

\begin{center}
\captionof{table}{Fornax type cluster: radius in [Mpc], baryonic mass in [$10^{12} M_\odot$], accelerations in [$10^{-10}$ m/s$^2]$}
\centering
\begin{tabular}{cccccccc}
\toprule
Radius & $M(< r)$ & $M(> r)$ & $a_N$ & $a_s$ & $a'_{0}$ & $a  \,(\ref{eq4})$ & $a /a_{\mathrm{MOND}}$ \\
\midrule
0.05 & 0.02 & 3.98 & 0.01 & 0.07 & 9.87 & 0.31 & 2.87 \\
0.10 & 0.09 & 3.91 & 0.01 & 0.05 & 6.24 & 0.28 & 2.28 \\
0.25 & 0.53 & 3.47 & 0.01 & 0.03 & 3.94 & 0.22 & 1.81 \\
0.50 & 1.80 & 2.20 & 0.01 & 0.01 & 2.56 & 0.16 & 1.46 \\
0.75 & 3.59 & 0.41 & 0.01 & 0.00 & 1.50 & 0.12 & 1.12 \\
0.80 & 4.00 & 0.01 & 0.01 & 0.00 & 1.20 & 0.10 & 1.00 \\
\bottomrule
\end{tabular}
\end{center}

\begin{center}
\centering
\includegraphics[scale=0.6]{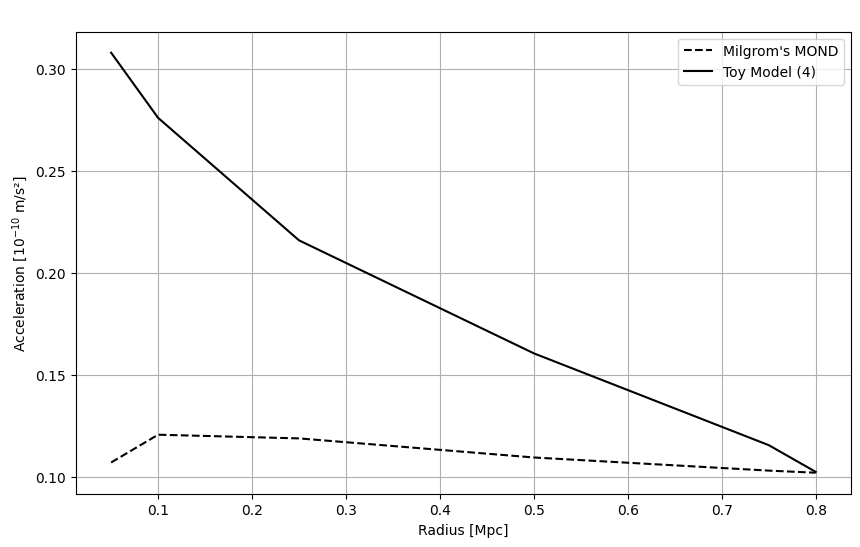}
\captionof{figure}{Plot for the Fornax cluster toy model showing the acceleration predicted by MOND $a_{\mathrm{MOND}} \, (\ref{eq1})$ and by the proposed $a \,(\ref{eq4})$. }
\label{Fig5}
\end{center}

\newpage

\begin{center}
\centering
\includegraphics[scale=0.6]{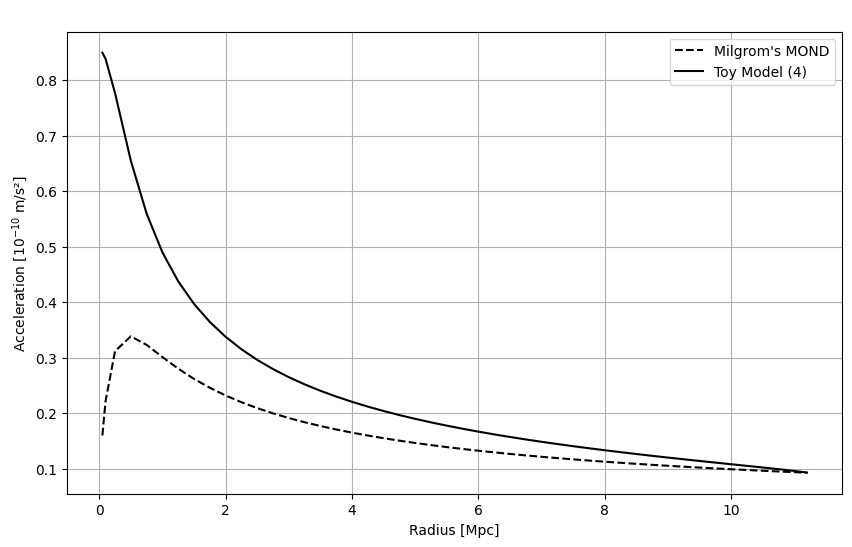}
\captionof{figure}{Plot for the Coma cluster toy model showing the acceleration predicted by MOND $a_{\mathrm{MOND}} \, (\ref{eq1})$ and by the proposed $a \,(\ref{eq4})$ with a virial radius of $4R$ as the upper bound of the integral (\ref{eq5}). }
\label{Fig6}
\end{center}

\end{appendices}

\end{document}